\documentclass[aps, amsfonts, nofootinbib, notitlepage, twocolumn, preprintnumbers]{revtex4-1}
\usepackage{hyperref}
\usepackage{ mathrsfs }
\hypersetup{
colorlinks=true,
linkcolor=red,
citecolor=blue,
urlcolor=blue}
\usepackage{epsf,epsfig}
\usepackage{amscd}
\usepackage{amsmath}
\usepackage{subfig}
\usepackage{graphicx}
\usepackage[countmax]{subfloat}
\input xy
\xyoption{all}

\newcommand{\bea}{\begin{eqnarray}}
\newcommand{\eea}{\end{eqnarray}}

\begin{document}

\title{Deconfinement Transition at High Isospin Chemical Potential and Low Temperature} 
\author{Thomas D.~Cohen\footnote{{\tt cohen@umd.edu}}} 
\author{Srimoyee Sen\footnote{{\tt srimoyee@umd.edu}}}
\affiliation{Maryland Center for Fundamental Physics,\\ 
Department of Physics,\\ 
University of Maryland, College Park, MD USA}

\preprint{UM-DOE/ER/...}

\begin{abstract}
We consider QCD with two degenerate flavors of  light quarks(up and down) at asymptotically high isospin $(\mu_I)$ with zero baryon chemical potential ($\mu_B$) and calculate for the first time a quantitative expression for the critical temperature of the deconfinement transition in this regime. At high isospin chemical potential and sufficiently low temperatures this theory becomes equivalent to a pure Yang-Mills theory and accordingly has a first order deconfinement phase transition. Although this was conjectured in a seminal paper by Son and Stephanov in the year $2001$, the critical temperature of this deconfinement phase transition was not computed. This paper computes the energy scale associated with this transition as a function of the chemical potential $\mu_I$ by relating the parameters of the equivalent Yang-Mills theory to those of the underlying theory. We also relate the equation of state in one strongly interacting regime of QCD namely at finite isospin density to that in pure Yang-Mills, with the latter being amenable to straightforward numerical calculation. Our results for the critical temperature of deconfinement transition can be compared with future lattice calculations.

\end{abstract}
\maketitle

\section{Introduction}
One of the goals of modern nuclear physics is to understand the rich structure of the QCD phase diagram at finite density and temperature. This is because various such phases of matter can be realized in nature under different circumstances. The core of neutron stars and the evolution of the early universe are two of the most prominent examples where a knowledge of the phase diagram would help our understanding.

It is very difficult to get a handle on this problem starting directly from QCD.  Accordingly it is useful to learn what we can from regimes which are tractable even when they are not of direct phenomenological relevance. In this paper we focus on the low temperature regime of  two-flavor QCD with zero baryon chemical potential and very large isospin chemical potential.  The study of this regime was originated by Son and Stephanov\cite{Son:2000xc,Son:2000by} who noted that in the limit of extreme isospin chemical potentials this theory became equivalent to a pure Yang-Mills theory. This implies among other things that unlike at zero chemical potential the theory has a first-order deconfining transition. Although, the first order deconfinement transition at high isospin chemical potential was conjectured by Son and Stephanov, the critical temperature for this deconfinement transition has not been calculated before. This paper, relates the properties of this emergent Yang-Mills theory to the parameters of the underlying theory and calculates the critical temperature of this first order deconfinement transition.

One exciting feature of this analysis is that we can relate the equation of state at large isospin chemical potential and low temperature to the equation of state of pure Yang-Mills theory. Since the equation of state of the pure Yang-Mills theory is computable using numerical studies on a lattice with relatively modest resources, this means that the equation of state of QCD at large isospin chemical potential and low temperatures is effectively computable with modest resources---including a computation of the critical temperature of the first order transition. 
\\\\
To put this problem in context, it is useful to recall that considerable progress has already been made in mapping the phase diagram both in isospin density vs temperature plane and baryon density vs temperature plane. The biggest hurdle impeding progress in this effort is the strength of the QCD coupling constant at moderate energy scales. A significant portion of the phase diagram which corresponds to observable phenomena such as the dynamics at the core of the neutron stars or the heavy ion collisions is inaccessible to perturbation theory calculations as the theory is strongly coupled in this regime. In order to circumvent this difficulty numerical calculations using lattice have been employed and this has been remarkably successful in some cases. One such success story was in cosmology, where it answered questions relating to whether the universe went through a phase transition while passing from a quark gluon plasma phase to a hadronic phase at temperatures around $150$ MeV. This problem was satisfactorily handled by finite temperature lattice QCD and it was concluded that the universe did not go through a QCD phase transition but a crossover as it cooled\cite{Karsch,Greensite20031}. But lattice QCD fails at finite baryon chemical potential at low temperature. The reason behind this is that the fermion determinant in the presence of a chemical potential for the quark number becomes complex, which causes lattice algorithms involving the method of important sampling to break down. This is known as the sign problem  \cite{Fodor:2001pe,Ejiri:2003dc,PhysRevD.71.114014,deForcrand:2008vr}. But an understanding of this regime of the phase diagram is absolutely necessary for neutron star physics. This is because the mass-radius relation and the transport properties of a neutron star are determined by the phase of matter at the core where baryon density is high and the temperatures are low \cite{Haensel,HATSUDA,Sagert,Sumiyoshi,Ohnishi2011284}. Efforts have been made to explore the finite baryon density regime using Nambu-Jona-Lasinio model \cite{PhysRev.122.345,Vogl:1991qt,RevModPhys.64.649,Hatsuda:1994pi,Buballa:2003qv}, Quark-Meson model \cite{Jungnickel:1995fp}, Polyakov loop extended Nambu-Jona-Lasinio model \cite{Meisinger:1995ih,Fukushima:2003fw,Ratti:2005jh,Roessner:2006xn,Sasaki:2006ww}, Polyakov loop extended quark-meson model \cite{PhysRevD.76.074023,PhysRevD.82.034029}. But these models are not very rigorous. Since the temperatures of interest are very low, and degrees of freedom involved are fermionic, Fermi liquid theory of nucleons, which is a theory of nucleon quasi particle excitaions about the Fermi surface is used to make predictions for baryon densities that correspond to a Fermi energy of $100$ MeV. Very high baryon density and low temperature regime on the other hand can be dealt with analytically. In this regime, it was conjectured that the quarks and the gluons became deconfined and the quarks at the Fermi surface were weakly coupled which made rigorous perturbative calculations possible \cite{Alford:2007xm}. The ground state of matter at highest baryon densities and low temperatures was found to be in a color-superconducting phase known as the color-flavor-locked phase where a condensate of Cooper pairs near the Fermi surface generated the color Meissner effect shielding the gluons.
\\\\
Apart from the regimes discussed in the previous paragraph, there exists another regime of the phase diagram, given by finite isospin chemical potential but zero baryon chemical potential. This paper explores the confinement dynamics in this regime. There are two reasons why this regime is of interest to us. The first reason is that in this regime lattice calculations are not plagued by the sign problem and hence may be practical. This means that lattice simulations can be used to test the results of model calculations and extrapolated results from rigorous perturbative calculations. In our paper we look at asymptotically high isospin chemical potential where we use perturbation theory to make predictions. Extrapolation of our results to moderate isospin densities, where perturbation theory breaks down can be compared with lattice results to determine the accuracy and the range of validity of the extrapolations. The second reason is that this regime may help give some insight into neutron stars as the imposition of charge neutrality in such objects(neutron star cores) leads to the suppression of proton fraction compared to the neutrons. This corresponds to a finite isospin chemical potential. Despite this, our system differs from the realistic scenario in these objects as there is also a finite baryon chemical potential in them as mentioned before. 

The intention of this paper is to look at the finite isospin regime for zero  $\mu_B$. This is largely because the intractability of a finite $\mu_B$ region using lattice simulations imply that our predictions for such a scenario cannot be tested for moderate densities. Also the introduction of a finite $\mu_B$ in the problem will further undermine the effect of the isospin chemical potential: the effect in which we are primarily interested. 
\\\\Note that our system is unstable with respect to weak decay and is not electrically neutral. Hence the thermodynamic limit of such a system does not exist. But we can ignore the electromagnetic and weak effects in order to isolate the strong dynamics \cite{Son:2000by}. There have been considerable recent developments in this region of the phase diagram \cite{Son:2000xc,Son:2000by,Kogut:2002zg,Kogut:2004zg,Sinclair:2006zm,deForcrand:2007uz,
Ebert:2005cs,Ebert:2009ty,Detmold:2012wc}.  

The phase structure of QCD at finite isospin chemical potential and zero baryon chemical potential with two light quarks was first considered in \cite{Son:2000by,Son:2000xc}. For small enough $\mu_I$, i.e. $\mu_I$ is much smaller than typical hadronic scales, chiral perturbation theory can be used to predict that charged pions($\pi^-$) start condensing for $\mu_I > m_{\pi}$ where $m_{\pi}$ is the mass of the pions. On the other hand, the opposite limit, where $\mu_I$ is much larger than typical hadronic scales and anti-up and down quarks form Fermi spheres of radius $+\frac{\mu_I}{2}$ respectively, interactions can be described using perturbative QCD due to asymptotic freedom. Just as in BCS in ordinary metals, if there is an instability at the Fermi surface caused by attractive interaction in some color, flavor and angular momentum channel, there will be a condensation of Cooper pairs in the problem at hand as well. In this high isospin regime, it is believed that, due to the strength of the attractive perturbative one gluon exchange interaction in the color singlet channel that a color singlet condensate of quark-antiquark Cooper pairs indeed does form. This means that two possible forms of the condensate are $\bar{u}\gamma_5 d$ and $\bar{u}d$. The instanton induced interaction favours the condensation in $\bar{u}\gamma_5 d$ channel over the $\bar{u}d$ one and hence the condensate has pion quantum numbers. It was conjectured that there was a smooth crossover between pion superfluid at low isospin to quark-antiquark condensate at high isospin density \cite{Son:2000by}. For an asymptotically high $\mu_I$ the phase diagram in $\mu_I-T$ plane was argued to have a first order deconfinement phase transition apart from the second order superconducting phase transition at a higher temperature. 

The argument is as follows. The condensate at high isospin density at temperatures smaller than the gap, has pion quantum numbers and is color neutral. This ensures that the gluons are not screened by the condensate. For temperatures well below the gap it is also energetically expensive to excite a quark quasi-particle out of the Fermi sphere and hence there is no Debye screening of the gluons either and we are left with a pure glue theory. A pure glue theory with $SU(3)$ gauge group is known \cite{Yaffe:1982qf} to have a first order deconfinement phase transition with increasing temperature. Therefore we can expect a similar behavior in our problem. It was argued by Son and Stephanov that the critical temperature of this deconfinement transition should be lower than the critical temperature of the superconducting transition \cite{Son:2000xc}. Interestingly, as there is no phase transition with increasing temperature for $\mu_I=0$, it was argued in \cite{Son:2000by} that the first order phase transition line has to end at a critical point if one assumes quark-hadron continuity along the $T=0$ isospin axis. 

 As mentioned above, It is expected that for $\Delta \gg T$ and $\mu_I \gg m_{\rho}$ the static color charges are not screened due to the absence of both Debye and Meissner masses and the theory describing the low energy dynamics of the system is governed by the Lagrangian density $\mathscr{L}=\frac{-F^2}{4g^2}$ where $g$ is the coupling constant that matches that of the original theory at the scale $\Delta$. But this is not a complete picture. Although the quarks are bound in $SU(3)$ singlet Cooper pairs, they can still partially screen the static color charges by changing the chromodielectric (dielectric constant for the color electric field) constant of the theory $\epsilon$ from unity. As we will see, in this theory $\epsilon >1$ and the coulomb potential between static color charges is smaller than $\frac{g^2}{r}$ by a factor of $\epsilon$ where $r$ is the distance between the static charges. This amounts to having a pure glue theory where the speed of gluon is given by $\frac{1}{\sqrt{\epsilon\lambda}} < 1$ where, $\lambda$ stands for the chromomagnetic permeability.   

While this general picture was introduced some time ago  \cite{Son:2000by} to the best of our knowledge, quantitative predictions of properties of the effective pure glue theory in terms of the underlying theory are lacking.  A principal purpose of this paper is to compute this chromodielectric constant (and the chromomagnetic permeability) as a function of $\mu_I$ integrating out the quarks around the Fermi surface in an energy range of the order of the gap. For asymptotically high $\mu_I$, we find that $\epsilon\lambda \gg 1$ where $\epsilon$ and $\lambda$ have been defined with respect to $\epsilon_0$ and $\lambda_0$ of the vacuum by absorbing the vacuum polarization effect in $g$. We also predict the effective confinement scale as a function of $\mu_I$. The confinement scale is related to the chromodielectric constant in a simple way which is also elaborated in this paper. The scaling of the confinement scale with $\mu_I$ obtained in this paper should be compared with scaling results obtained from lattice simulations if possible in future.  

A particularly interesting fact is that there is a first-order deconfinement transition at large $\mu_I$.  We cannot analytically compute its critical temperature from first principles. However,if we can determine the ratio of the deconfinement transition temperature to the scale of the the theory for pure Yang-Mills, we can then predict the phase transition temperature for large $\mu_I$.  Moreover, as noted artier, the calculation of the deconfinement transition temperature for pure Yang-Mills is  computable on the lattice with quite modest resources, so that a prediction of the QCD deconfinement phase transition at large $\mu_I$ is viable as is a more general calculation of the equation of state.  The equation of state turns out to be different from the equation of state of pure Yang Mills theory suitably rescaled.  Apart from the low lying gluoynamics, the theory has a meson Goldstone mode which contributes to the equation of state.  While the mode is massless, and thus it contributes at low energies, the scale of its interactions is well above scale of the gluodynamics. Thus, while the Goldstone mode contributes to the pressure, it is also effectively non-interacting and does not affect the gluopynamics.

The story of a pure glue theory anisotropic in space-time at sufficiently high isospin chemical potential is reminiscent of the `two flavor color-superconductor' ($2SC$) phase at low temperature and high $\mu_B$ \cite{Rischke:2000cn}. In case of the $2SC$ phase at finite baryon density, two quarks form a Cooper pair giving rise to a condensate that breaks color $SU(3)$ down to color $SU(2)$. Five of the eight gluons become massive due to the color-Meissner effect of the condensate, but the remaining three remain massless. At temperatures lower compared to the gap Debye screening for these three gluons is absent giving rise to pure $SU(2)$ gluodynamics. However, just like in our problem, the quark-loops alter the chromodielectric constant of the system. 

The paper is organized as follows: Sec. \ref{micro} discusses the microscopic Lagrangian that is used to determine the low energy constants of the gluodynamics. Sec. \ref{effaction} outlines the evaluation of the chromodielectric constant and the chromomagnetic permeability using the microscopic Lagrangian. Sec. \ref{confinement} estimates the energy scale of the deconfinement transition.  Sec. \ref{EOS} discusses the connection of the equation of state of QCD at large isospin chemical potential and low temperature to that of pure Yang Mills and computes the critical temperature for the deconfinement transition followed by a concluding section.

\section{Microscopic Lagrangian}\label{micro}
In order to quantify the effect of the paired quarks on the chromodielectric constant of the system we need to be able to write down the effective theory for the gluons. The low energy constants of this theory need to be extracted by integrating out quark loops from a microscopic Lagrangian, which in this case is the QCD Lagrangian. We emphasize again that since we are working at asymptotically high densities it is justified to make use of perturbation theory. We start with the QCD Lagrangian with two light (massless) quarks in the fundamental representation of the $SU(3)$ color group with an isospin chemical potential $\mu_I$,

\bea
\mathscr{L}=\overline{\psi}(i\gamma^\mu D_\mu+\mu_I \gamma^0 \tau_3)\psi-\frac{1}{4}(F^a_{\mu\nu})^2
\eea
where,
\bea
\psi=\begin{pmatrix}
u \\
d
\end{pmatrix}
\label{micro}
\eea , $F^a_{\mu\nu}=\partial_\mu A^a_{\nu}-\partial_{\nu}A^a_{\mu}+g f_{abc}A^b_{\mu}A^c_\nu$ and $D_{\mu}=\partial_{\mu}-igA^a_{\mu}t^a$. The generators of color $SU(3)$ are denoted by $t^a$ and $\tau_3$ is the third Pauli matrix given by 
\bea
\tau_3=
\frac{1}{2}\begin{pmatrix}
1 && 0\\
0 && -1
\end{pmatrix}.
\eea $u$ and $d$ stand for up and down quarks respectively. As mentioned above the condensate has pion quantum numbers and is of the form $\bar{u}\gamma_5d$. In order to analyse this condensate we define
\bea
\tilde{\psi}\equiv\begin{pmatrix}
u \\
\tilde{d}
\end{pmatrix}
\eea where $\tilde{d}=\gamma_5 d$. We intend to find the propagator for the $\tilde{\psi}$ quarks in the presence of this condensate and to do so we introduce an auxiliary field $\Delta$ such that the inverse propagator for the quarks is given by 
\bea
G^{-1}=-i\begin{pmatrix}
\gamma^{\mu}p_{\mu}+\mu_I \gamma^0 && \Delta \\
\tilde{\Delta} && \gamma^{\mu}p_{\mu}-\mu_I \gamma^0
\end{pmatrix}
\eea with $\tilde{\Delta}=\gamma_0\Delta^{\dagger}\gamma_0$. The quarks are expected to acquire a gap in their spectrum which is to say that it is energetically expensive to excite a quark bound in a Cooper pair. This gap is characterized by the magnitude of the auxiliary field when the action is minimized with respect to the auxiliary field. This amounts to solving for the gap, $\Delta$ using Schwinger-Dyson (S-D) equation. We do not need to solve for $\Delta$ for the purpose of our paper, althouh it is straightforward to do so. The gap was estimated in \cite{Son:2000by} to be of the form $\Delta=b|\mu_I|g^{-5}\exp(-c/g)$ where $c=\frac{3\pi^2}{2}$ and $g$ was evaluated at the scale $\mu_I$. Note that, in ordinary superconductivity with a four-fermi interaction the gap is of the form $\Delta\sim \exp(-k/g^2)$ where $k$ is some numerical constant. Here, however, $\Delta\sim \exp(-c/g)$. The source of this behavior is that instead of a contact four-fermi interaction we have long range magnetic interaction, just as in the case of a superconducting gap at large $\mu_B$. $b$ was estimated to be $\sim 10^4$ \cite{Son:2000by}. It is easy to see from the expression for the gap that as $g$ gets smaller the gap $\Delta$ becomes smaller compared to $\mu_I$. This is a standard feature of any BCS calculation and is the only consistent solution of the gap equation other than $\Delta=0$. In our problem as $\mu_I$ becomes large, the strength of the coupling becomes smaller and we obtain $\mu_I \gg \Delta$.
\\\\
For the sake of clarity we explore the Dirac structure of the condensate here to some extent. Forgetting for a moment that the condensate is a constant in space-time, we write it as a function of momentum as $\Delta(k)$. The gap matrix $\tilde{\Delta}(k)$ is diagonal in color space but not in Dirac space. Just like any $4\times 4$ matrix in the Dirac space it can be expanded in the basis of $16$ linearly independent $4\times 4$ matrices. We however, are interested in the channel with a total angular momentum of zero. If we restrict ourselves to the $J=0$ channel, the basis includes only $8$ Lorentz invariant matrices, $1, \gamma_{\mu}k^{\mu}, \gamma_{\mu}u^{\mu}, \gamma_{\mu}k^{\mu}\gamma_{\nu}u^{\nu}, \gamma_5, \gamma_5\gamma_{\mu}k^{\mu}, \gamma_5\gamma_{\mu}u^{\mu}, \gamma_5\gamma_{\mu}k^{\mu}\gamma_{\nu}u^{\nu}$ where $u^{\mu}$ is the $4$ velocity of the medium. To restrict ourselves more, we are interested only in the odd parity channel. This is because, as mentioned before, although perturbative one gluon exchange gives rise to equally strong attractive interaction in both $\bar{u}d$ (even parity) and $\bar{u}\gamma_5d$ (odd parity) channels, the instanton induced interaction favors the $\bar{u}\gamma_5d$ channel. The requirement of odd parity shrinks the basis to $1, \gamma_{\mu}k^{\mu},  \gamma_{\mu}u^{\mu}, \gamma_{\mu}k^{\mu}\gamma_{\nu}u^{\nu}$. More over, it can be expected that at asymptotically high isospin densities, chiral symmetry is restored asymptotically. The evidence of this comes from \cite{Son:2000by}. There it is shown that for $\mu_I >m_\pi$, the value of the chiral condensate is given by $<\bar{u}u+\bar{d}d> =2\bar{\psi}\psi \cos(\alpha)$ and that of the charged pion condensate is given by $<\bar{u}\gamma^5 d> +\text{h.c.} =2\bar{\psi}\psi \sin(\alpha)$, where, $\cos(\alpha)=\frac{m_\pi^2}{\mu_I^2}$.  There we notice that as the isospin chemical potential reaches $m_{\pi}$ from zero and grows further, the condensate starts rotating from a chirally broken $\bar{q}q$ phase to a chirally symmetric charged pion phase. The condensate is expected to crossover to a completely chirally symmetric phase at high $\mu_I$. The imposition of chiral symmetry on the condensate gets rid of $\gamma_{\mu}k^{\mu}$ and $\gamma_{\mu}u^{\mu}$ from the basis. We can finally express our order parameter in the reference frame of the medium i.e. $u=(1,0)$ as 
\bea
\tilde{\Delta}=i\Delta_4+i\Delta_5\hat{\gamma} .\hat{k}\gamma_0 \\
=i\left(\Delta^{+}\lambda^{+}_{\mathbf{p}}+\Delta^{-}\lambda^{-}_{\mathbf{p}}\right)
\label{Dtilde1}
\eea
.
$\lambda^{+}_{\mathbf{p}}$ and $\lambda^{-}_{\mathbf{p}}$ are free quark on-shell projectors given by
\bea
\lambda^{\pm}_{\mathbf{p}} \equiv \frac{1}{2}\left(1\pm \gamma_0 \boldsymbol{\gamma}.\boldsymbol{\hat{p}}\right)
\eea,
where $\Delta^{+}$ and $\Delta^{-}$ are real numbers. 
If we restrict ourselves to the $^1S_{0}$ channel, we have $\Delta^{+}=\Delta^{-}$. For the purpose of this paper we are only interested in the $^1S_0$ channel and we do take $\Delta^{+}=\Delta^{-}=\Delta$. With this structure of the condensate in mind, we return to deriving the propagator for the $\tilde{\psi}$ quarks. We do so by going to Euclidean time which corresponds to a finite temperature calculation and then take the limit $T\rightarrow 0$ at the end. 
The inverse quark-propagator in Euclidean action is given by
\bea
G^{-1}=\begin{pmatrix}
-i\gamma^0 p^0-\gamma^i p^i +\mu\gamma^0 && \Delta \\
\tilde{\Delta} && -i\gamma^0 p^0-\gamma^i p^i -\mu\gamma^0 
\end{pmatrix}
\eea.
Inverting $G^{-1}$ we arrive at the propagator
\bea
G=\begin{pmatrix}
G^{+} && \Sigma^{-}\\
\Sigma^{+} && G^{-}
\end{pmatrix}
\eea
where
\begin{align}
\begin{split}
G^{-}\left(k\right)=\sum_{e=\pm}\frac{-ik_0+(\mu-e k)}{-k_0^2-(\epsilon^e_k)^2}\lambda^{-e}_{\mathbf{k}}\gamma^0 \\
G^{+}\left(k\right)=\sum_{e=\pm}\frac{-ik_0-(\mu-e k)}{-k_0^2-(\epsilon^e_k)^2}\lambda^{+e}_{\mathbf{k}}\gamma^0 \\
\Sigma^{-}(k)=i\sum_{e=\pm}\frac{\Delta^e \lambda^e}{-k_0^2-(\epsilon^e_k)^2}\\
\Sigma^{+}(k)=-i\sum_{e=\pm}\frac{\Delta^e \lambda^{-e}}{-k_0^2-(\epsilon^e_k)^2}
\end{split}
\label{G}
\end{align}
and $(\epsilon^e_k)^2=(\lvert\mathbf{k}\rvert-e \mu)^2+(\Delta^{e})^2$ where $e=\pm$. Note that, as expected, the dispersion relation of the quarks have become gapped. Also, the dispersion relations with $e=-1$ correspond to anti-up and down quarks where as the ones with $+1$ correspond to up and anti-down quarks. Apart from the propagator of the quarks in this basis, that is $\tilde{\psi}$ quarks, we also need the interactions of the $\tilde{\psi}$ quarks with the gluons to make any progress. The interaction term between the quarks and the gluons in the QCD Lagrangian, i. e. $\bar{\psi}\gamma^{\mu}A^{a}_{\mu}t^a\psi$ needs to be rewritten in the basis of the $\tilde{\psi}$ quarks. For the sake of clarity we write down explicitly the interaction term between the gauge field and $\tilde{\psi}$ quarks and it is given by
\bea
\bar{\tilde{\psi}}\begin{pmatrix}
-ig\gamma^\mu A^a_{\mu}t^a && 0 \\
0 && -ig\gamma^\mu A^a_{\mu}t^a
\end{pmatrix}\tilde{\psi}
\eea.
\section{effective action}\label{effaction}
Armed with the propagator for the quark and the interaction of the quarks with the gluons we can now write down the form of the low energy effective Lagrangian from symmetry arguments alone. The microscopic Lagrangian had $SU(3)_{color}\times SU(2)_{flavor}\times U(1)_{em}$ internal symmetries, none of which are broken by the condensate. This means that the low energy effective theory should respect these internal symmetries. The microscopic Lagrangian however, was not Lorentz invariant due to the presence of the chemical potential which broke Lorentz symmetry to rotational invariance explicitly. This means that the effective Lagrangian for the gluons will be rotationally invariant but not Lorentz invariant. Apart from this, by requiring locality the form of the effective action can be obtained as
\bea
S_{glue}=\sum_{a}\frac{1}{g^2}\int d^4q\left(\frac{\epsilon}{2}E_a^2-\frac{1}{2\lambda}B_{a}^2\right)\left(1+O\left(\frac{q^2}{\Delta^2}\right)\right)
\label{eff}\eea, where $E^a$ and $B^a$ are chromoelectric and chromomagnetic fields respectively, $g$ is the QCD coupling at energy scale $\mu_I$. For gluon momentum $q$, smaller than the gap $\Delta$, higher dimensional operators which include powers of $q$ or more powers of $E$ and $B$ fields are supressed by powers of $\frac{|q|}{\Delta}$ and the effective action behaves like a pure Yang-Mills theory. $\epsilon$ and $\lambda$ are the low energy constants, to be identified as the chromodielectric constant and the chromomagnetic permeability of the medium. These can be explicitly calculated by computing the polarization tensor for the gluons, which amounts to integrating out the quark quasiparticles upto energies of the order of the gap around the Fermi surface. In other words, we use the one loop polarization tensor to obtain the part of the effective action that is quadratic in the gluon fields and also contains the effects of the integrated out quarks. In this action for the gluons, we rewrite the gluon fields in terms of the chromoelectric and chromomagnetic fields using the definitions 
\bea
E^a_i=F^a_{0i}
\eea and
\bea
B^a_k=\epsilon_{ijk}F_{ij}
\eea and compare with Eq. \ref{eff}. The two Lagrangians are compared in order to write down $\epsilon$ and $\lambda$ in terms of the various components of the polarization. Upto leading order in the expansion of gluon momentum over the gap, at zero temperature this matching calculation yields the following relations:
\begin{align}
\begin{split}
\Pi^{00}_{ab}(q_0,\mathbf{q})=-(\epsilon -1)|\mathbf{q}|^2\delta_{ab}\\
\Pi^{ij}_{ab}\frac{\delta^{ij}}{2}\frac{\delta_{ab}}{8}=-(\epsilon -1)q_0^2\\
\Pi^{ij}_{ab}(q_0,\mathbf{q})\left(-\delta^{ik}-\frac{q^iq^k}{|\mathbf{q}|^2}\right)=\left(\frac{1}{\lambda}-1\right)|\mathbf{q}|^2\delta_{ab}\delta^{jk}
\end{split}
\label{P}
\end{align}
where, $\Pi$ is the polarization of the gluons. The indices in superscripts on $\Pi$ stand for space-time indices and the subscript for color indices.
\\
We now have to calculate the polarization tensor. At one loop it is given by
\begin{align}
\begin{split}
\Pi^{ij}_{ab}(p)&=g^2T\sum_{k_0}\int \frac{d^3k}{(2\pi)^3}\text{Trace}\left[\gamma^i t_a G^{+}(k)\gamma^j t_b
G^{+}(k-p)\right.\\
&\qquad\quad \left. +\gamma^i t_a G^{-}(k)\gamma^j t_b 
G^{-}(k-p)\right.\\
&\qquad\quad \left. +\gamma^i t_a \Sigma^{-}(k)\gamma^j t_b
\Sigma^{+}(k-p)\right.\\
&\qquad\quad \left. +\gamma^i t_a \Sigma^{+}(k)\gamma^j t_b
\Sigma^{-}(k-p)\right]
\end{split}\label{Pi}\end{align}. The expressions for $G^{+}, G^{-}, \Sigma^{+}$ and $\Sigma^{-}$ are substituted from eq. \ref{G} in eq. \ref{Pi}. It is easy to see after the substitution that $\Pi^{00}_{11}=\Pi^{00}_{22}=\Pi^{00}_{33}$ and so on. 
\\
At this point we should note that the trace in eq. \ref{Pi} is a standard one and especially similar to the unbroken $SU(2)$ gluon polarization in the two flavor color-superconducting phase. The reason behind this similarity has been explained before and is related to the fact that the low energy dynamics of both the phases are driven by gluodynamics. The above mentioned trace was calculated by \cite{Rischke:2000qz} and the integral over 3 momentum was computed in \cite{Rischke:2000cn} in the context of the $2SC$ phase. We should remember that the three momentum integral is dominated by anti-up and down quarks with momenta close to the Fermi surface. Hence, the up and anti-down quark contribution to the above integral is neglected which amounts to neglecting terms with denominators $(\lvert\mathbf{k}\rvert-e \mu)^2+(\Delta^{e})^2$ or $(\lvert(\mathbf{k}-\mathbf{p})\rvert-e \mu)^2+(\Delta^{e})^2$with $e=-1$. Because the trace and the integral in eq. \ref{Pi} are standard, instead of going into the details of the calculation we just state the results here. We find that as $T\rightarrow 0$, around $q_0,\mathbf{q}=0$ up to leading order $\frac{|q|^2}{(\Delta^{+})^2}$ expansion, $\Pi^{00}_{aa}=-\frac{g^2\mu_I^2|\mathbf{q}|^2}{18\pi^2(\Delta^{+})^2}$ and $\Pi^{ij}_{ab}=\frac{g^2\mu_I^2q_0^2}{18\pi^2(\Delta^{+})^2}\delta^{ij}\delta_{ab}$. The appearance of $\mu_I$ in the numerator is due to the fact that the integral in eq. \ref{Pi} is dominated by momenta near the Fermi surface. This means that the three momentum integral essentially turns into a one dimensional integral multiplied by the area of the Fermi surface which is proportional to the square of the Fermi momentum or the chemical potential. Note that $\Pi^{ij}_{ab}(-\delta^{jk}-\frac{q^iq^k}{|\mathbf{q}|^2})=0$. Comparing eq. \ref{P} with these expressions for $\Pi^{00}_{aa}$, $\Pi^{ij}_{ab}$ we conclude that $\lambda=1$ and $\epsilon=1+\frac{g^2\mu_I^2}{18\pi^2(\Delta^{+})^2}$. 
\section{confinement}\label{confinement}

\begin{figure}
\includegraphics[width=.4\textwidth]{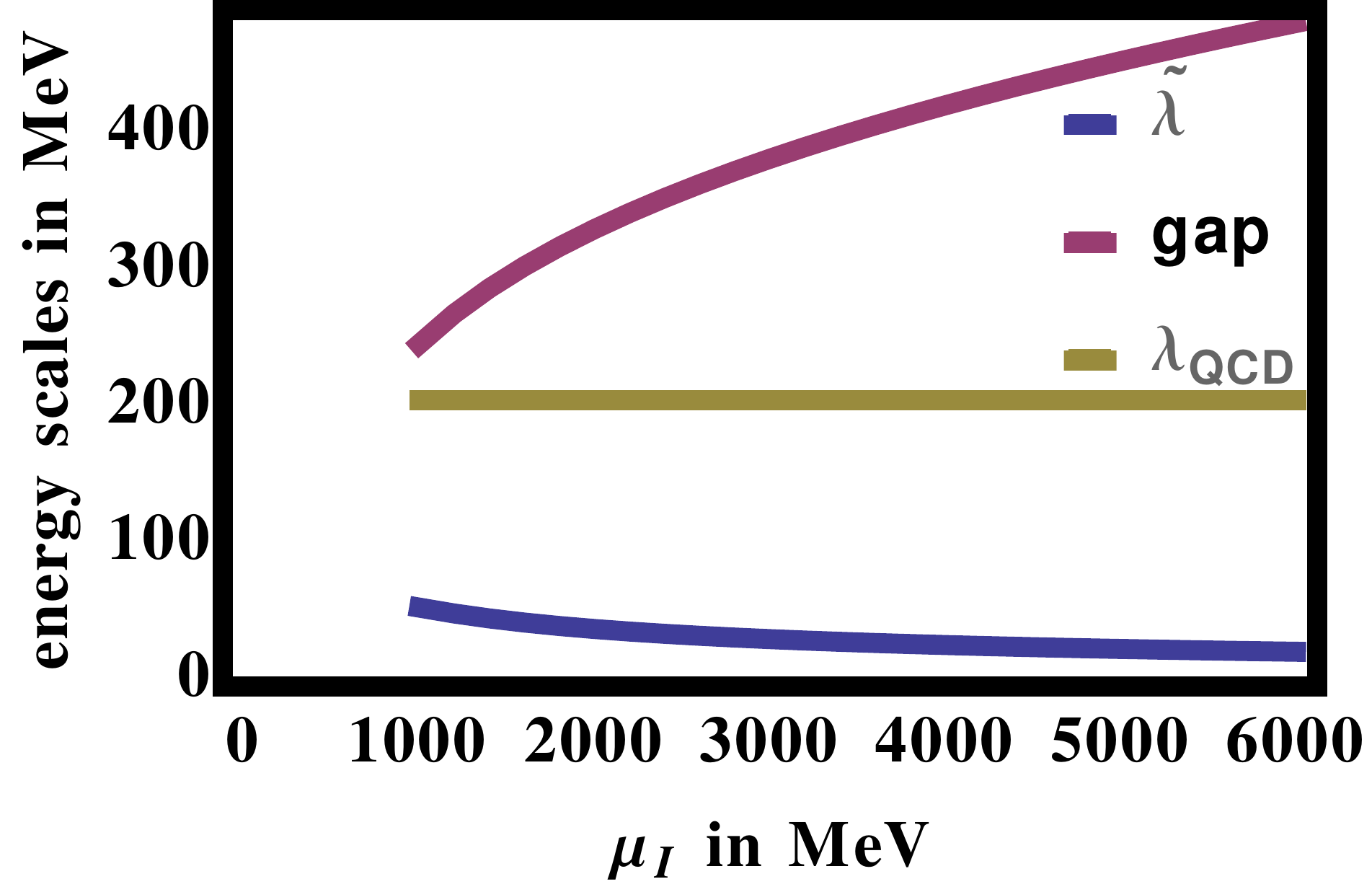}

\caption{
Confinement scale in MeV as a function of chemical potential in MeV
}
\label{f2}
\end{figure}

\begin{figure}
\includegraphics[width=.4\textwidth]{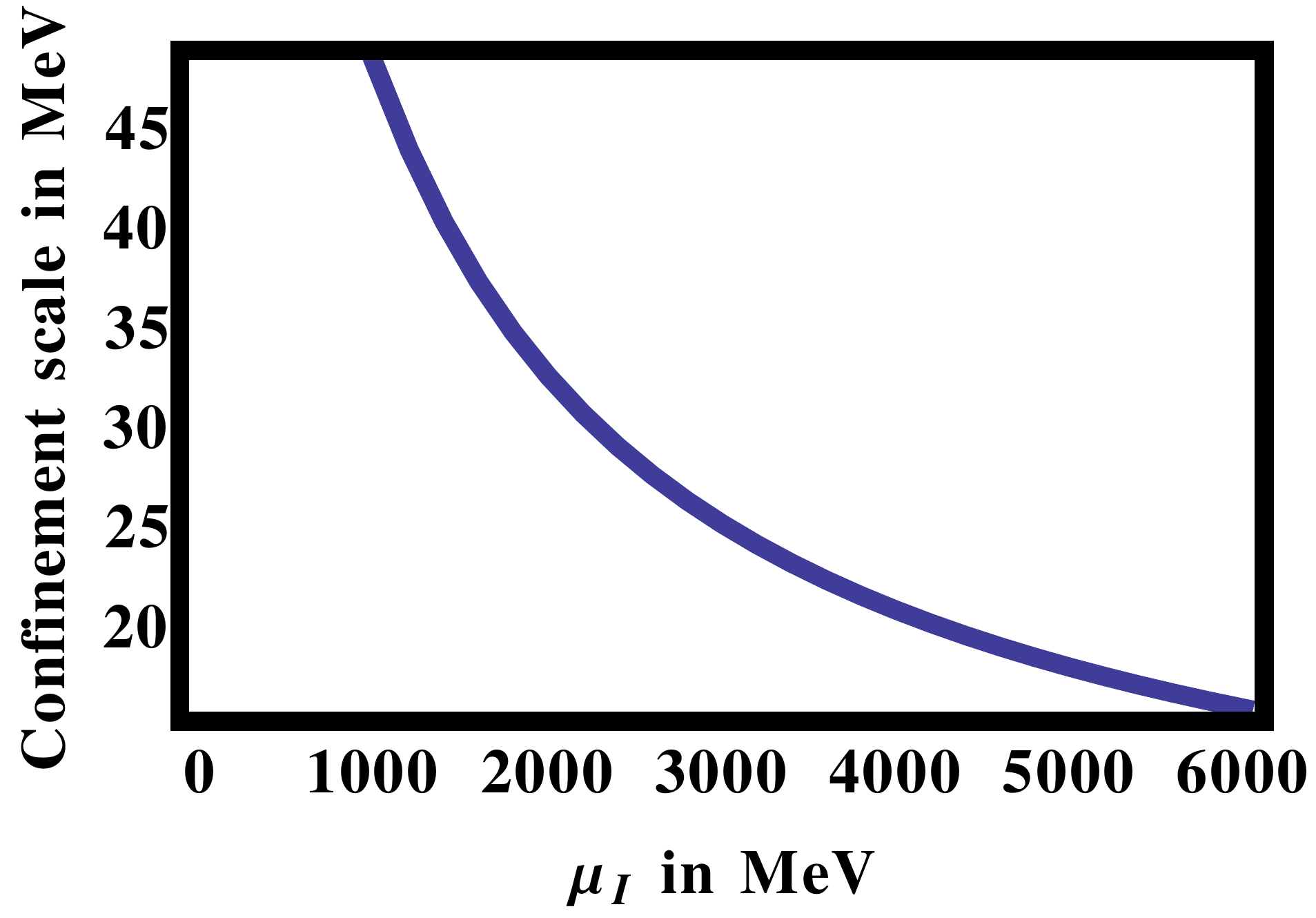}

\caption{
The QCD confinement scale, the confinement scale at high isospin and the gap are shown in the same plot. The regime towards the right with higher isospin is where all these scales are well separated and we trust our calculations. The scales get closer as we go towards the left of the plot where the isospin chemical potential is small. This is the regime where a lattice calculation could determine how accurate the predictions are.
}
\label{f1}
\end{figure}

The effective Lagrangian that we obtained by integrating out the fermions has the form of a pure glue theory upto corrections $\sim \frac{|q|}{\Delta}$ and can be recast as 
\bea
S=-\frac{1}{4(g')^2}\int d^4x' (F')^2
\label{eff2}
\eea
with some simple rescalings of the fields and the coupling etc. The action of \ref{eff2} will however have a speed of gluon given by 
\bea
c=\frac{1}{\sqrt{\lambda\epsilon}}=\frac{1}{\sqrt{\epsilon}}\\
\approx \frac{3\sqrt{2}\pi\Delta}{g\mu_I} 
\eea as $\lambda=1$ and for $\mu_I\gg\Delta$.
 Before elaborating on the scaling, we would like to point out that $g'$, in eq. \ref{eff2} runs exactly like the coupling in a pure $SU(3)$ Yang-Mills theory for momentums smaller than the gap. As the energy scales reach the gap from below and become bigger than the gap, the running behavior of the coupling will no longer be that of pure $SU(3)$ Yang-Mills and will get affected by the higher dimensional operators. Hence, the action of eq. \ref{eff2} should be matched with the actions of eq. \ref{eff} at the scale $\Delta$. To know how exactly this matching is done, we have to rescale time, fields and the coupling in eq. \ref{eff} as follows. We can rescale time direction, the gauge field and the coupling as $t_0''=\frac{t_0}{\sqrt{\epsilon}}$, $A_0^{a''}=\sqrt{\epsilon}A_0^a$ and $g''=\frac{g}{\epsilon^{1/4}}$ to rewrite the Lagrangian in eq. \ref{eff} as 
 \bea
S=-\frac{1}{4(g'')^2}\int d^4x'' (F'')^2
\label{eff3}
\eea where $
\alpha_s''=\frac{\alpha_s(\mu)}{\sqrt{\epsilon}}\ll \alpha_s
$. Now we have to match the coupling constant in the action \ref{eff2}, $\alpha_s'(q')$ with $\alpha_s''$ at a scale $q'=\Delta$ which in turn means
\bea
\alpha_s'(\Delta)=\alpha_s''=\frac{\alpha_s(\mu)}{\sqrt{\epsilon}}\ll \alpha_s \; .
\eea
With this pure Yang-Mills theory in hand, we now concentrate on the dynamics of the deconfinement transition. As mentioned in the introduction, a pure $SU(3)$ Yang-Mills theory undergoes a first order deconfinement phase transition with increasing temperature. This means that in our problem at finite temperature, for temperatures smaller compared to the gap there will be a deconfinement phase transition with increasing temperature. The energy scale of this deconfinement transition will be much smaller than the gap and we call this new confinement scale $\tilde{\lambda}$ to distinguish it from the zero density QCD confinement scale $\lambda_{QCD}$. We can figure out how the confinement scale of the effective theory compares with the gap by looking at the renormalization group equation
\begin{align}
\begin{split}
\tilde{\lambda}&=\Delta\exp\left(-\frac{2\pi}{b_0\alpha_s'(\Delta)}\right)\\
&=\text{$\Delta$} \text{Exp}\left[-2\sqrt{2\pi }\frac{\mu }{29\text{$\Delta $}\sqrt{\alpha_s'(\Delta) }}\right]
\label{L}
\end{split}
\end{align}
where $b_0$ at one loop with two flavours for color $SU(3)$ is given by $b_0=11-\frac{2}{3}n_f=\frac{29}{3}$ and $\alpha_s'$ is the coupling evaluated at the scale $\Delta$. To evaluate the confinement scale we have to evaluate the gap. The dependence of $\Delta$ on the coupling is
\bea
\Delta^{+}=b|\mu_I |g(\mu_I)^{-5}\exp(-\frac{3\pi^2}{2g(\mu_I)})
\eea
with $b=10^4$ \citep{Son:2000xc}. We compute and plot the confinement scale for a range of the isospin chemical potential from asymptotically large values where we trust our calculations completely to regions of smaller values where the approximations start breaking down. In the region of smaller isospin density the confinement scale $\tilde{\lambda}$ could be appreciably different from what we found here. It would be instructive to compare our results with a future lattice calculation in the regime of small isospin density $\leq 1000$ MeV where the different scales in the theory are not so well separated. Fig. \ref{f2} illustrates the various energy scales in the problem as a function of $\mu_I$. It also demonstrates how the scales get widely separated for asymptotically high isospin density where perturbation theory is valid and how they get closer in the opposite limit with low isospin density where the QCD coupling becomes considerably strong. For example if we look at fig. \ref{f2} for $\mu_I \sim 5000$ MeV, we notice that $\Delta \sim 400$ MeV and this is an indication of the fact that our calculations are trustworthy in this regime as the gap is much smaller than the isospin chemical potential. As expected from the eq. \ref{L} the confinement scale decreases exponentially with the isospin chemical potential and this is illustrated in fig. \ref{f1}.


\section{Equation of State and the critical temperature of deconfinement transition}\label{EOS}
As mentioned in the introduction, it is possible to use our analysis to extract more information about the system than have already computed above.  The key additional information is the equation of state as a function of temperature in the regime of large $\mu_I$ and $T \ll \Delta$ including the value of  the critical temperature of the deconfinement transition.  Of course,  It is well known that even for pure Yang-Mills theory the equation of state and the critical temperature are not computable analytically due to the strength of the coupling.  However, it can be computed on lattice using numerical techniques, and the lattice calculations to do are far simpler than for theories including dynamical quarks. If we parametrize the relation between the  transition  for pure Yang-Mills with the confinement scale of the theory in the following way
\bea
T^c_{YM}=\kappa_{YM}\lambda_{YM}
\label{kappa}
\eea
where $T^c_{YM}$ is the critical temperature for the deconfinement transition, $\lambda_{YM}$ is the confinement scale for pure Yang-Mills, $\kappa_{YM}$, the proportionality constant between the two can be computed using lattice. The form of the relation between the critical temperature and confinement scale in Eq. \ref{kappa} follows from dimensional analysis alone.  

We have already computed the confinement scale($\tilde{\lambda}$) for the low energy effective theory for our problem with QCD with two light quarks and a very high isospin chemical potential as a function of $\lambda_{QCD}$, the QCD confinement scale at zero density. 
Thus the critical temperature $\tilde{T_c}$ in our problem, will be related to $\tilde{\lambda}$ as 
\bea
\tilde{T_c}=\kappa_{YM}\tilde{\lambda}.
\label{kappa2}
\eea

The equation of state of QCD in the high isospin chemical potential limit with two light quark flavors and $T \ll \Delta$ can also be related to that of pure Yang-Mills theory. In order to do this we need to discuss one additional aspect of the problem in detail. Other than the unscreened gluons at low energy scales there is also a Goldstone mode associated with the spontaneous breaking of isospin rotations about the $3$ direction in isospin space.
The Goldstone mode corresponds to a rotation of 
\bea
\begin{pmatrix}
u\\
d
\end{pmatrix}\rightarrow e^{i\alpha\tau_3}\begin{pmatrix}
u\\
d
\end{pmatrix}.
\eea
The microscopic Lagrangian of Eq. \ref{micro} is invariant under this transformation but the condensate $\bar{u}\gamma_5 d$ is not.  Clearly this Goldstone mode contributes to the equation of state. However, it is also clear that the Goldstone mode does not interact with the dynamics of the  gluons below the energy scale of the gap as the interactions are suppressed by powers of $\mu_I$. Hence the deconfinement phase transition in the gluodynamics is not affected by the presence of this Goldstone mode. However, while the Goldstone mode does not interact with the glueballs,  it does contribute to the pressure as a function of temperature---that is the equation of state, and in a formal sense,  the contribution from the Goldstone mode is of the same order in temperature as that of the gluons. Fortunately, as the gluons and the Goldstone mode do not interact and the two contributions can be computed separately and added to give the pressure. The contribution to the pressure from the Goldstone mode is given by
\bea
P_{\text{Goldstone}}=-\int \frac{d^3p}{(2\pi)^3}\text{ln}(1-e^{\frac{vp}{T}})=\frac{\pi^2}{90}\frac{T^4}{v^4}
\eea
where $T$ is the temperature, $v$ is the speed of the Goldstone mode in the medium and is given by $1/\sqrt{3}$. This number $1/\sqrt{3}$ is a result that is true for Goldstone modes in any finite density system \cite{Beane:2000ms}. The contribution coming from the gluons to the pressure is identical in form to that in pure Yang-Mills and can be computed on lattice. This contribution can be parametrized as 
\bea
P_{\text{gluon}}=T^4 f\left(\frac{T}{\tilde{T_c}}\right)
\label{Pglue}
\eea
where $f$ is a function of the temperature and the critical temperature. Here, note that the speed of gluons in the rescaled theory is given by $1$. Again the form of the pressure in Eq. \ref{Pglue} can be obtained by dimensional analysis. The functional form of $f\left(\frac{T}{T^c_{YM}}\right)$ as a function of $T$ and $T^c_{YM}$ can be computed using lattice. Hence we conclude that the equation of state for QCD with two light flavors of quark at a very high isospin chemical potential is given by
\bea
P(T)=T^4 f\left(\frac{T}{\tilde{T_c}}\right)+\frac{\pi^2}{90}\frac{T^4}{v^4}.
\eea
where $f$ is a numerically calculable function of temperature.

\section{Conclusion}\label{sec:conclusion}
The phenomenon of a deconfinement transition with increasing temperature at a very high isospin density might seem surprising in that a similar phenomenon is not expected at high baryon chemical potential ($\mu_B$). Given its unfamiliarity, it is useful to try to put in context what our results tell us about the nature of the phase diagram in the $T-\mu_I$ plane at zero $\mu_B$. Our calculations allowing the determination of the equation of state are only reliable in the high $\mu_I$ and low $T$ region. Nevertheless, one can deduce a few qualitative features about the phase diagram.
\begin{figure}

\begin{tabular}{ccc}
\subfloat[]
{\label{fig:f0}
\includegraphics[width=0.4\textwidth]{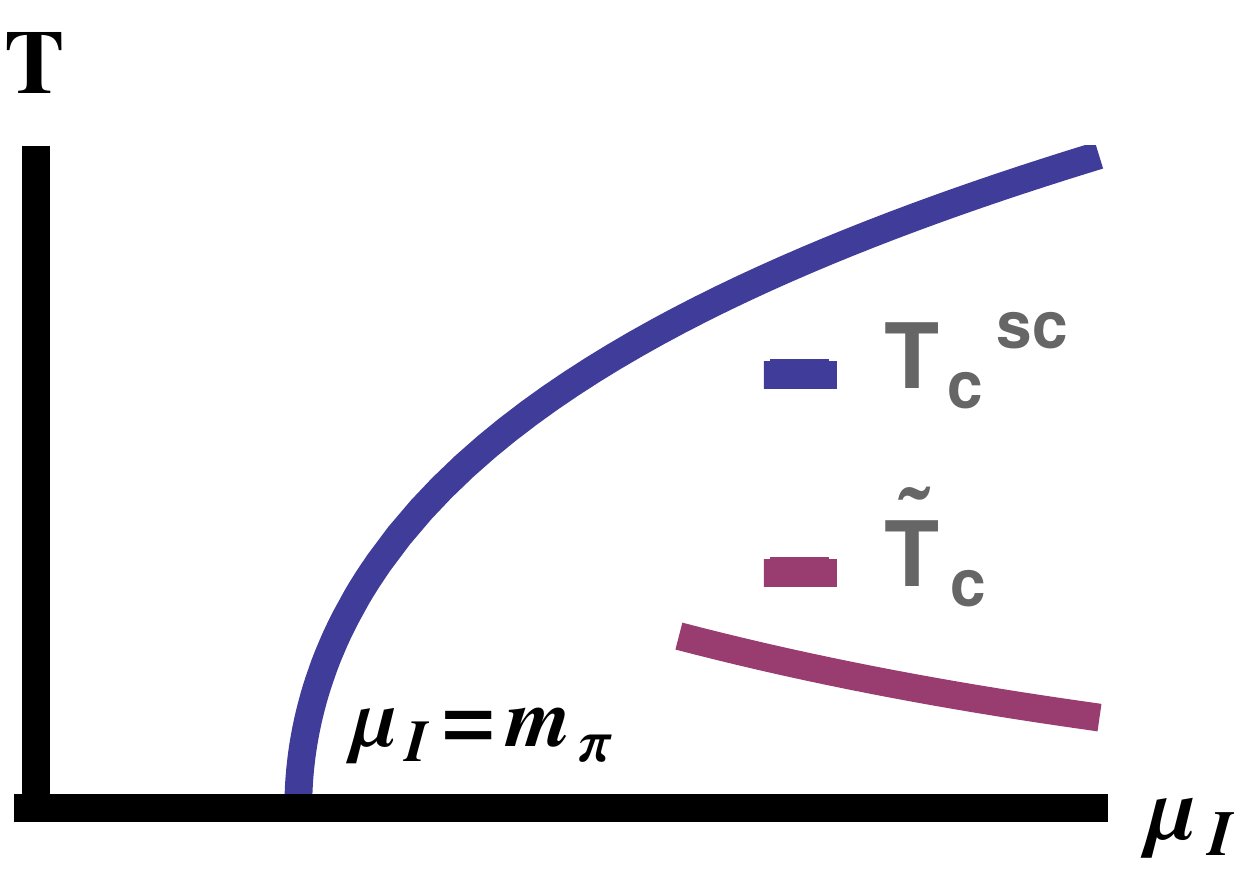}}
\\
\subfloat[]
{\label{fig:f1}
\includegraphics[width=0.4\textwidth]{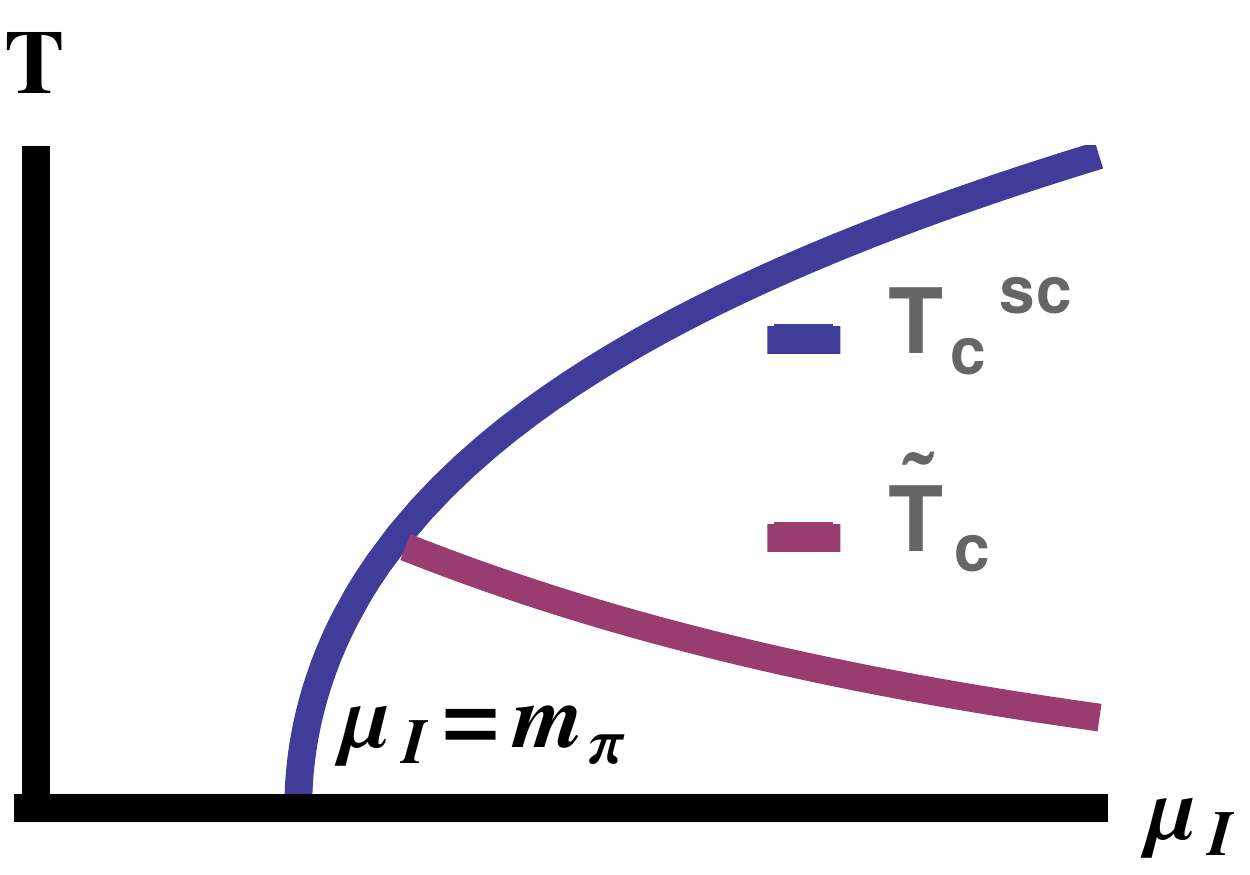}}
\\
\subfloat[]
{\label{fig:f2}
\includegraphics[width=0.4\textwidth]{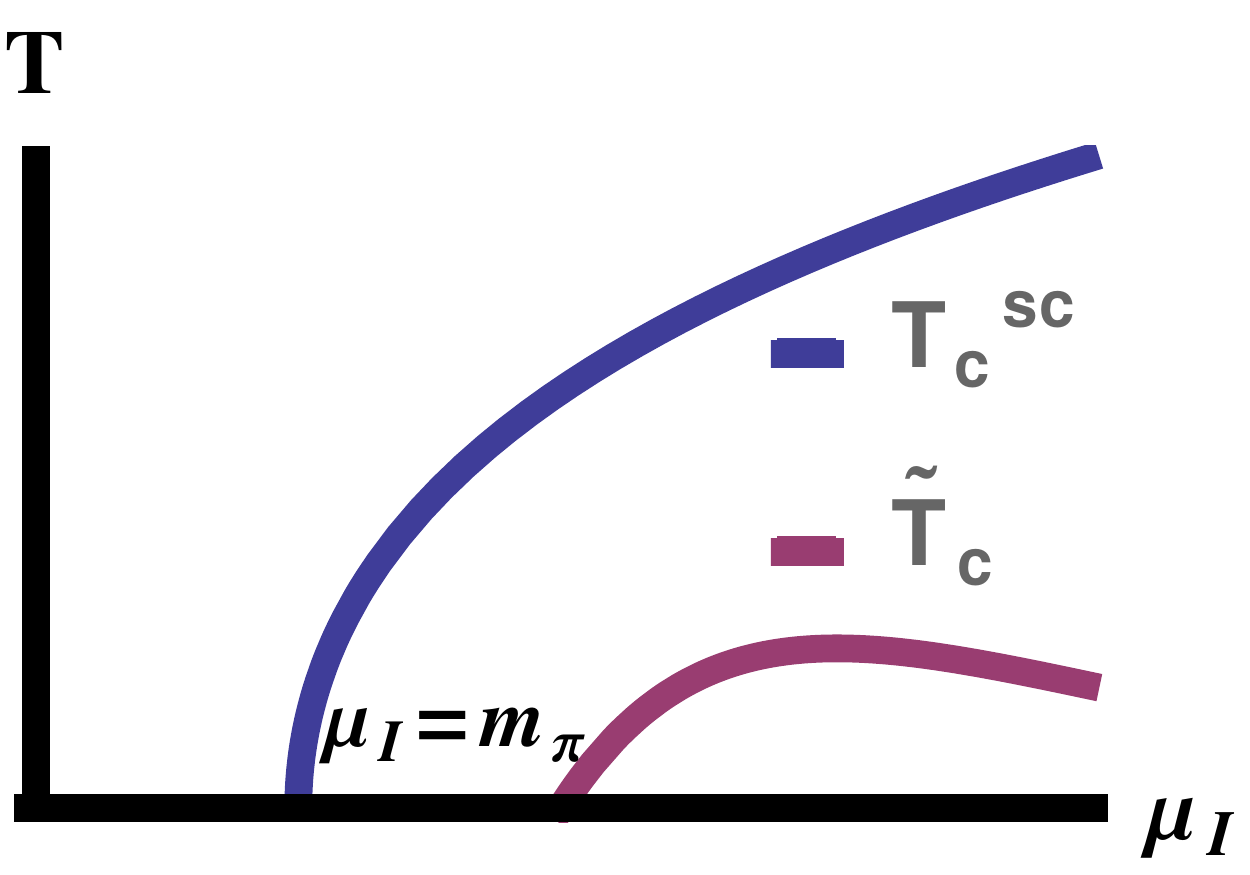}}
\end{tabular}
\caption{\label{fig:multi_panel} 
The three possibile scenarios describing what happens to the deconfinement phase transition line as $\mu_I$ is lowered. $T_c^{sc}$ corresponds to the critical temperature for the superconducting transition and $\tilde{T_c}$ is the critical temperature for the deconfinement transition.}\label{fig:conc}
\end{figure} 
Several features of the phase diagram should be regarded as being on solid ground. The first is that at $\mu_I =0$, there are no phase transitions as $T$ changes. The second is that the for sufficiently high $\mu_I$ there exists a superconducting phase with a gap serving as an order parameter. At low temperatures and isospin densities this superconducting phase may be regarded as a pion condensate. At high temperature however this condensate melts. Thus there must exist a curve which separates the superconducting phase from the normal phase. The third robust feature of the phase diagram was the focus of this paper---namely the existence of a first-order phase transition at sufficiently high $\mu_I$ between a confining and a deconfining phase. This implies the existence of a curve which separates the confined phase from the deconfined phase. The critical temperature of the deconfinement transition is proportional to the confinement scale we have evaluated above as written in Eq. \ref{kappa2}. As was seen in fig. \ref{fig:f2} it decreases exponentially with increasing isospin density in the regime where we trust our calculations. But in the regime where our calculations are not valid the behavior of this line is a matter of speculation. 

These three facts imply the existence of certain special points in the phase diagram.  For the moment let us assume that regardless of $\mu_I$, there are at most two phase transitions as a function of $T$, one associated with superconductivity and the other associated with confinement.  However,  as
 was pointed out in \cite{Son:2000xc} we also know that curves associated with these transitions must end somewhere at sufficiently low $\mu_I$, since there are no transitions at $\mu_I=0$. The points where the curves end are special. For the case of the superconducting transition, this special point occurs at $T=0$ and $\mu_I=m_\pi$ as is seen in all three curves in Fig. \ref{fig:conc}. An interesting and nontrivial fact is that one can infer the existence of an additional special point where the deconfinement transition ends.  Son and Stephanov  speculated about the nature of this point\cite{Son:2000xc}.  They argued that it is very plausible that there is no phase transition at zero $T$ as one increases $\mu_I$ from $m_\pi$ (where the superconducting transition occurs) as there is no symmetry.  A very natural way for this to occur is the one suggested in \cite{Son:2000xc} and shown in Fig. \ref{fig:f0}, namely for the curve of first-order transitions separating the confined from deconfined phases to end at a critical point. This possibility is particularly exciting since critical points are characterized by universal behavior.  However, we should note that it is not guaranteed to occur. For example, one could imagine that the special point associated with the end of the confinement transition could be a triple point as shown in Fig. \ref{fig:f1}  Alternatively, one might imagine, that despite the plausibility argument in ref.~\cite{Son:2000xc}, that the curve might end at $T=0$ as in Fig. \ref{fig:f2}  indicating a second transition at $T=0$.    One can also imagine more baroque scenarios with additional phases.  The point, though is that however, the curve ends, it must end somewhere and the point at which this happens is intrinsically of interest. Ultimately, one would hope that lattice calculations can determine the phase structure and the position of these interesting points.

\section*{Acknowledgements}
S. S. would like to thank Naoki Yamamoto for insightful discussions and suggestions. This work was supported by the U.S. Department of Energy through grant number DEFG02-93ER-40762.

\bibliographystyle{unsrt}
\bibliography{isospin}
\end{document}